\documentclass[a4paper, 12pt]{paper}
\usepackage{graphicx}        
\usepackage{cite}
\usepackage[warn]{mathtext} 
\usepackage{amsmath}
\usepackage{amssymb}
\usepackage{amsfonts}
\usepackage[unicode]{hyperref}
\usepackage[left=2cm,right=2cm, top=2cm,bottom=2cm,bindingoffset=0cm]{geometry}

\begin{document}
\title{Mechanism of solitary state appearance in an ensemble of nonlocally coupled Lozi maps}

\author{Nadezhda Semenova\thanks{{nadya.i.semenova@gmail.com}}, \and Tatyana Vadivasova, \and 
Vadim Anishchenko }
%
%
%
%
%
%
%
\maketitle

\textbf{Abstract}
We study the peculiarities of the solitary state appearance in the ensemble of nonlocally coupled chaotic maps. We show that nonlocal coupling and features of the partial elements lead to arising of multistability in the system. The existence of solitary state is caused by formation of two attractive sets with different basins of attraction. Their basins are analysed depending on coupling parameters.

\section*{Introduction}\label{sec:intro}

One of an inexhaustible area of research is connected with nonlinear ensembles with a large number of elements and different network topologies, which give rise to various types of spatio-temporal dynamics. Thus, besides widely considered chimera states (see, for example, \cite{KUR02a, ABR04, OME11, OME13, ZAK14, PAN15, JAR15, SEM15a, SEM17a, DAI18, GOP18, GHO18}), ensembles with nonlocal coupling can demonstrate another, recently found spatio-temporal structure, which is called ``solitary state'' \cite{MAI14a, JAR15, SEM15a, SEM17a, HIZ16a,HIZ16b, SEM15b, PRE16, PRE17,JAR18,GOP18, SAT18, SHE17_nd}. In contrast to the chimera state (which consists of spatially divided clusters of coherent and incoherent behaviour), the solitary state regime is characterized by a coherent behaviour of the whole system, except several elements. These elements do not form a cluster and for this reason these oscillators are called solitary ones. Their characteristics, conditions for the appearance and bifurcation mechanisms are still unexplored still. One of the possible mechanisms of solitary state emergence has been described in \cite{JAR18} for an ensemble of  phase oscillators with inertia. It has been shown that solitary states arise in a homoclinic bifurcation of a saddle-type synchronized state and die eventually in a crisis bifurcation after essential variation of the parameters. However, the solitary state appearance probably depends on a type of individual elements and a type of their interaction. For this reason we suppose that this mechanism is not a universal one. There are only a few works in this direction in contrast to the structures of chimera states. In the other hand, this effect is apparently typical for the dynamics of real multicomponent networks such as neuron ensembles, computer and power grids, and etc.

Solitary states can be observed in ensembles with global and almost global coupling \cite{MAI14a, HIZ16a, SHE17_nd}, nonlocal coupling \cite{JAR15, SEM15a, SEM17a}, and in some cases with local coupling \cite{JAR18}. Solitary states can coexist with chimera states. For example, the paper \cite{JAR15} shows that the chimera state can be formed through the solitary state regime. In split of various types of partial elements, there is a class of elements for which the chimera state can be observed in ensembles with nonlocal coupling and the solitary state is not \cite{SEM15a}. This class includes chaotic maps and chaotic time-continuous oscillators which are characterized by the transition to chaos via period-doubling bifurcations (Feugenbaum scenario). The transition from spatial coherence to incoherence in such ensembles occurs through chimera states. Peculiarities of their formation and their characteristics have been described in a number of works (see, for example, \cite{OME11, OME12, SEM15a, SEM17a, BOG17}). Some of them \cite{SEM15a, SEM17a} show that the type of a chaotic attractor impacts on the appearance of different spatial structures. For example, if we consider a ring of nonlocally coupled H{\'e}non maps (they have a nonhyperbolic chaotic attractor) this system can demonstrate chimera states. But if we replace H{\'e}non maps with Lozi maps, that exhibit quasihyperbolic attractors, such a system would not show chimera states and the transition to spatial incoherence would occur through solitary states \cite{SEM15a, SEM17a}.

Therefore, the question arises: ``What is the bifurcation mechanism of solitary state formation in an ensemble of elements such as the Lozi map?'' In the present paper we investigate an ensemble of nonlocally coupled Lozi maps and analyse in detail the dynamics of partial elements, which leads to the appearance of solitary states. We show that the nonlocal coupling can give rise to bistability in the ensemble, and this is one of the reasons of the emergence of solitary states formation.

\section{System under study} \label{sec:system}
Let us consider the ensemble of coupled maps:
\begin{equation}\label{eq:system}
\begin{array}{l}
x^{t+1}_i=f_x(x^t_i,y^t_i)  +  \frac{\sigma}{2P}\sum\limits^{i+P}_{i-P}  \large[ f_x(x^t_j,y^t_j) - f_x(x^t_i,y^t_i) \large], \\ 
y^{t+1}_i=f_y(x^t_i,y^t_i), \ \ \ \ \ \ \ i=1,2,\dots N.
\end{array}
\end{equation}
Here $f_x(x^t_i,y^t_i)$ and $f_y(x^t_i,y^t_i)$ are the functions which are defined by right-hand parts of the two-dimensional Lozi map, $x_i$, $y_i$ are the state variables, $t$ is the discrete time, $N$ is the number of elements in the ensemble. The nonlocal coupling is characterized by the coupling strength $\sigma$ and the number of neighbours $2P$ ($P$ neighbours on the either side of the $i$th element). For simplification we add the coupling only to the first equation (\ref{eq:system}). It is not essential for our investigations.

It has been shown in \cite{SEM15a, SEM17a} that the ensemble dynamics depends on individual elements. In the present work we are focused on the system (\ref{eq:system}) with Lozi maps:
\begin{equation} \label{eq:Lozi}
x^{t+1}=f_x(x^t, y^t)=1-\alpha|x^t|+y^t, \ \ \ \ \ \ \ \ \ \ \ y^{t+1}=f_y(x^t,y^t)=\beta x^t,
\end{equation}
where we fix the parameters $\alpha$ and $\beta$ which correspond to a chaotic regime in an individual element. In this case one can obtain solitary states during the transition from complete chaotic synchronization to spatial incoherence in a certain ranges of coupling parameters $\sigma$ and $r=P/N$. These states have already been found in works \cite{SEM15a, SEM17a}, however, no explanation has been given there.


\section{Numerical analysis of formation and evolution of solitary states in the ensemble of Lozi maps} \label{sec:SS_formation}

Varying of the coupling parameters leads to a sequence of spatio-temporal regimes with the appearance of solitary states. Their evolution has been described in \cite{SEM15a, SEM17a} without revealing of the bifurcation mechanism. To understand the mechanism of solitary state appearance, let us consider the results of numerical simulation for the transition from coherence to incoherence in more detail.

\begin{figure}[htbp]
\centering
\resizebox{0.8\columnwidth}{!}{\includegraphics{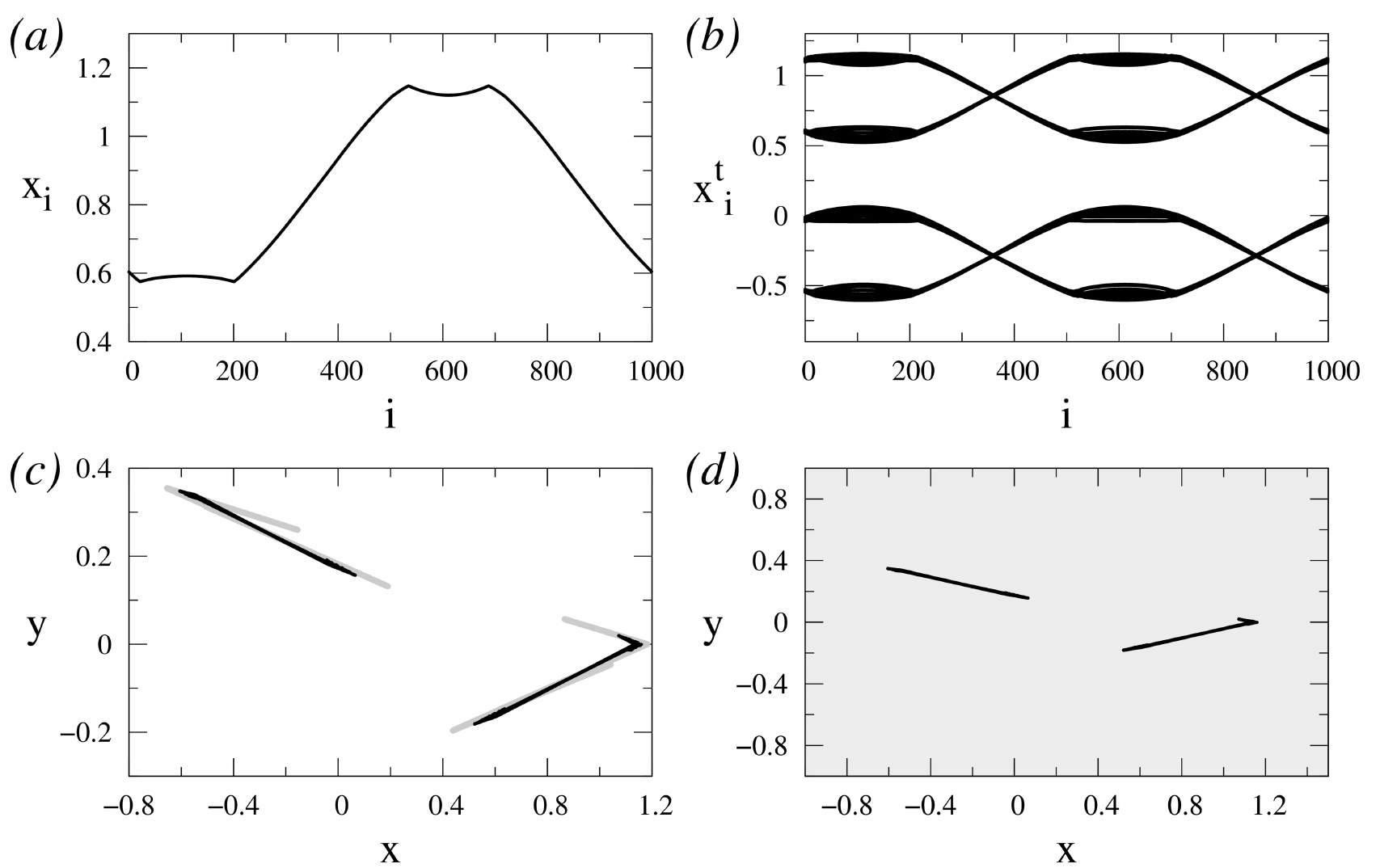} }
\caption{Spatial coherence in the ensemble (\ref{eq:system}) of Lozi maps for $\sigma=0.27$. Top panels illustrate the instantaneous snapshot at the time $t=10000$ (a) and the last 50 instantaneous spatial profiles $x^t_i$ (b). Panel (c) depicts the attracting set for the ensemble (\ref{eq:system}) on the phase plane $(x,y)$ (black dots) and the attractor on an isolated Lozi map (\ref{eq:Lozi}) (gray points). The set for the ensemble (\ref{eq:system}) and its basin of attraction are shown in panel (d) for the oscillator $k=600$. Other parameters: $\alpha=1.4$, $\beta=0.3$, $r=0.2$.}
\label{fig:L_coherence}
\end{figure}

We fix $\alpha=1.4$ and $\beta=0.3$ corresponding to a chaotic regime in the Lozi map (\ref{eq:Lozi}). In this case the attractor consists of two parts. On the next step we fix the coupling radius $r=0.2$ and consider a stable regime for $\sigma=0.27$ and random initial conditions ($x^0_i$, $y^0_i$ are randomly distributed in the intervals $x^0_i\in[-0.5;0.5]$ and $y^0_i\in[-0.6;0.6]$). It corresponds to spatial coherence with continuous instantaneous spatial profiles in the ensemble (\ref{eq:system}) of Lozi maps \cite{SEM15a, SEM17a} (Fig.~\ref{fig:L_coherence}). It means that $|x_i-x_{i+1}|<\delta$, $\delta\ll 0$ for neighbouring oscillators. The panel (a) of Fig.~\ref{eq:system} shows an instantaneous spatial profile (snapshot) at a certain time. The panel (b) corresponds to the last $50$ spatial profiles. This illustration is called a spatio-temporal profile \cite{BOG17}. It can be seen that all the profiles are distributed near four curves. It indicates that the dynamics is almost periodic. To illustrate the corresponding set, let us imagine these profiles $(x^t_i$, $y^t_i)$ in the phase plane $(x,y)$. The corresponding plot is shown in Fig.~\ref{fig:L_coherence}c, where we compare this set (black points) and the set demonstrated by one uncoupled Lozi map (\ref{eq:Lozi}) for $\alpha=1.4$ and $\beta=0.3$ (gray points). As can be seen, these sets are almost identical.

The distinctive feature of the ensemble (\ref{eq:system}) of Lozi maps is that all points of the instantaneous spatial profile belong to one band of the double-band attractor. It takes place when the coupling parameters are varied. However, it does not occur in the ensemble (\ref{eq:system}) of H{\'e}non maps (see \cite{SEM15a, SEM17a}). Thus, we can assume that this distinction is caused by different transitions to chaos in the H{\'e}non and Lozi maps, as it has been mentioned in Section \ref{sec:intro}. 

The attractor in Fig.~\ref{fig:L_coherence}d is a single attracting set for one oscillator. The basin of attraction for the $k$th oscillator can be computed as follows. We use the initial conditions randomly distributed in the intervals $x^0_i \in [-0.5;0.5]$, $y^0_i \in [-0.6;0.6]$, $i=1,2,\dots N$. After $5\cdot 10^4$ iterations a steady state appears in the system. Then we change the initial conditions of the $k$th oscillator in the intervals $x^0_k \in [-1.5;2]$ and $y^0_k \in [-1;1.5]$ and calculate the basins of attraction for different sets. For $\sigma=0.27$, all trajectories from different initial conditions $x^0_k$, $y^0_k$ for the $k$th oscillator reach the same set (Fig.~\ref{fig:L_coherence}d). The same results have been obtained for the other oscillators.

Now let us consider the impact of the coupling strength. The initial conditions are the same as in the previous coherence regime ($\sigma=0.27$). When $\sigma$ increases, the system (\ref{eq:system}) shows the transition from coherence to complete chaotic synchronization, which is not so important in our studies. Let us consider main regimes which can be realized when decreasing $\sigma$.

\begin{figure}[htbp]
\centering
\resizebox{0.8\columnwidth}{!}{\includegraphics{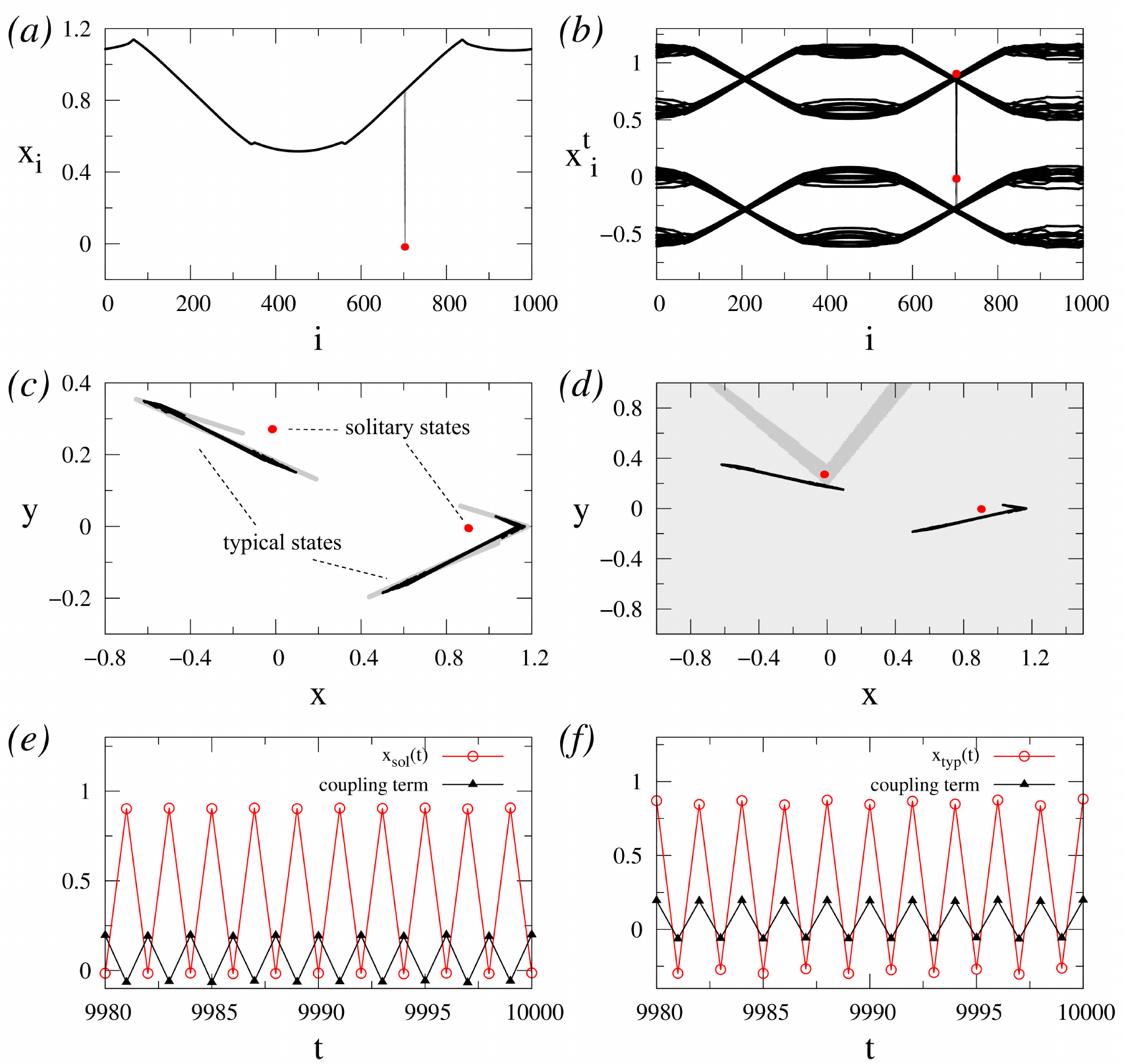} }
\caption{Solitary state in the ensemble (\ref{eq:system}) of Lozi maps for $\sigma=0.226$. Top panels illustrate the instantaneous snapshot at $t=10000$ (a) and the last 50 instantaneous spatial profiles $x^t_i$ (b). Panel (c) depicts the attracting sets for the solitary oscillator (red points) and for the other oscillators (black dots) of the ensemble (\ref{eq:system}) in the phase plane $(x,y)$ and the attractor in an isolated Lozi map (\ref{eq:Lozi}) (gray points). The sets for the ensemble (\ref{eq:system}) and their basins of attraction are shown in panel (d) for the oscillator $k=702$. The light-gray area corresponds to the basin of attraction for the typical set; the gray region is the same but for the solitary set. Panels (e) and (f) show the realizations of the system (\ref{eq:system}) for solitary oscillator ($i=702$ in (e)) and typical oscillator ($i=701$ in (f)) and the corresponding coupling terms $\frac{\sigma}{2P} \times \sum\limits^{i+P}_{j=i-P} f(x^t_j,y^t_j),\ \ j\neq i$.}
\label{fig:L_SS1}
\end{figure}

When $\sigma=0.226$, the first solitary oscillator $i=702$ appears in the ensemble (\ref{eq:system}). The other oscillators are in the coherence regime. The instantaneous snapshot (panel (a)) and the last $50$ instantaneous spatial profiles (panel (b)) are shown in Fig.~\ref{fig:L_SS1}a,b. The spatio-temporal profiles illustrate the period-2 dynamics of the solitary oscillator, which is anti-phase with the other oscillators. Now there are two sets on the phase plane $(x,y)$ (Fig.~\ref{fig:L_SS1}c). The first set (black points) is the same set as in the coherence regime, and the second set (red dots) corresponds to the solitary oscillator with periodic dynamics. For convenient description, let us refer to the first set as a typical set and the second one as a solitary set.

Figure \ref{fig:L_SS1}d depicts the basins of attraction for the typical and solitary sets prepared for the oscillator $k=702$ in the case when the other oscillators belong to the left top set. It can be seen that the solitary set has the basin of attraction in the form of V-letter. This domain of non-zero measure is much smaller than another basin. The same basins can be obtained for the other oscillators. It means that using specially prepared initial conditions one can obtain several solitary states for the same parameters. Nevertheless, the basin is rather narrow. Therefore, the solitary oscillator is only one for random initial conditions. 

\begin{figure}[htbp]
\centering
\resizebox{0.8\columnwidth}{!}{\includegraphics{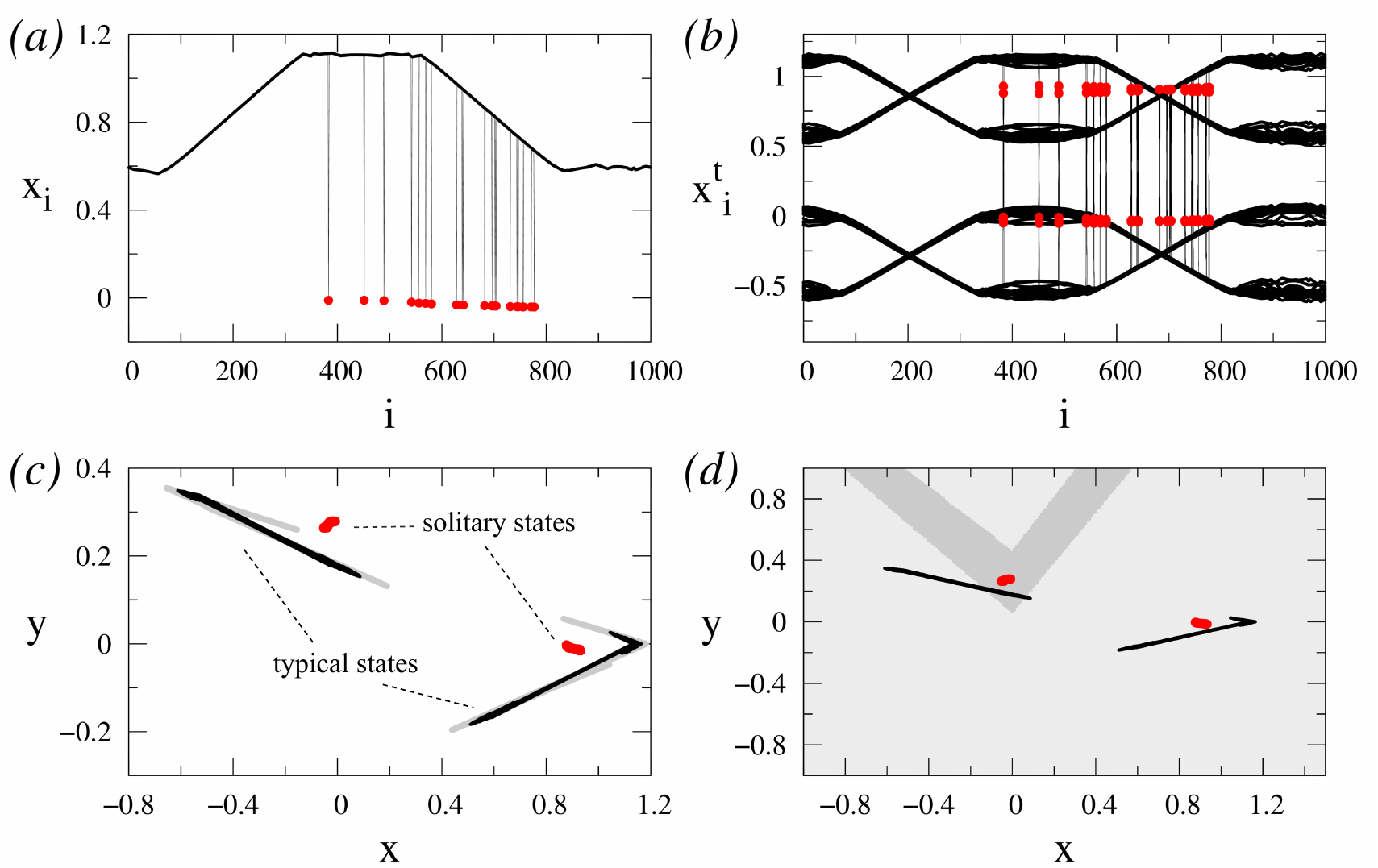} }
\caption{Solitary state regime in the ensemble (\ref{eq:system}) of Lozi maps for $\sigma=0.22$. Top panels show the instantaneous snapshot at $t=10000$ (a) and the last 50 instantaneous spatial profiles $x^t_i$ (b). Panel (c) displays the attracting sets for the solitary oscillators (red points) and for the other oscillators (black dots) of the ensemble (\ref{eq:system}) on the phase plane $(x,y)$ and the attractor in an isolated Lozi map (\ref{eq:Lozi}) (gray points). The sets for the ensemble (\ref{eq:system}) and their basins of attraction are shown in panel (d) for the oscillator $k=580$. The light-gray area corresponds to the basin of attraction for the typical set; the gray region is the same but for the solitary set. Other parameters: $\alpha=1.4$, $\beta=0.3$, $r=0.2$.}
\label{fig:L_SS_many}
\end{figure}

When decreasing of the coupling strength $\sigma$, the number of solitary oscillators increases (Fig.~\ref{fig:L_SS_many}a,b). They start influencing on the solitary states and the dynamics of the whole ensemble. For this reason the solitary set involves new points and becomes more blurred (Fig.~\ref{fig:L_SS_many}). The recorded growth leads to an enlargement of the corresponding basin of attraction (Fig.~\ref{fig:L_SS_many}). This finding explains a growing number of solitary oscillators obtained from initial conditions.

\begin{figure}[htbp]
\centering
\resizebox{0.8\columnwidth}{!}{\includegraphics{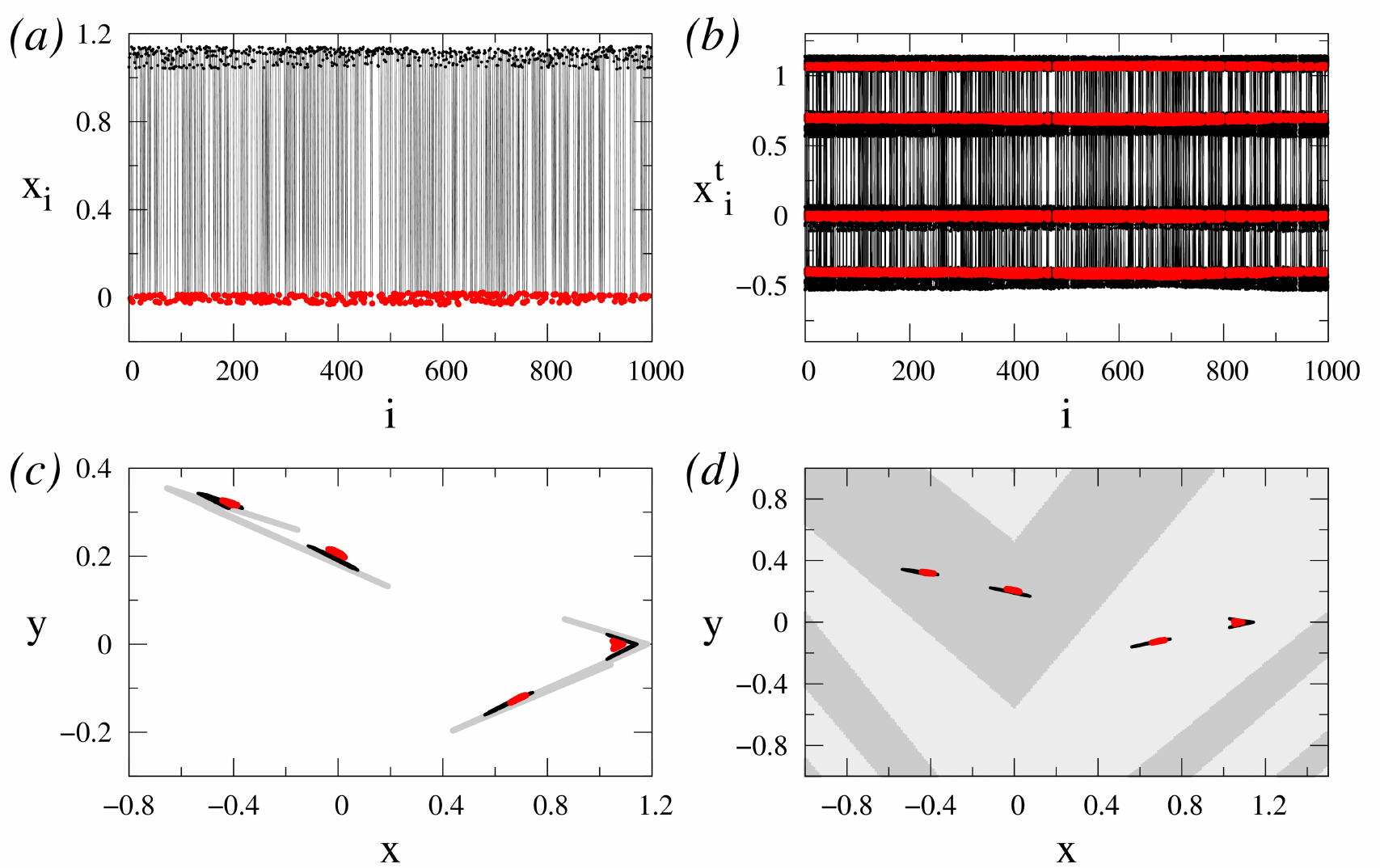} }
\caption{Transition to spatio-temporal chaos in the ensemble (\ref{eq:system}) of Lozi maps (\ref{eq:Lozi}) for $\sigma=0.1$. Top panels reflect the instantaneous snapshot at $t=10000$ (a) and the last 50 instantaneous spatial profiles $x^t_i$ (b). Panel (c) shows the attracting sets for the solitary oscillators (red points) and the other oscillators (black dots) of the ensemble (\ref{eq:system}) in the phase plane $(x,y)$ and the attractor in an isolated Lozi map (\ref{eq:Lozi}) (gray points). The sets for the ensemble (\ref{eq:system}) and their basins of attraction are plotted in panel (d) for the oscillator $k=350$. The light-gray area corresponds to the basin of attraction for the typical set; the gray region is the same but for the solitary set. Other parameters: $\alpha=1.4$, $\beta=0.3$ and $r=0.2$.}
\label{fig:L_incoherence}
\end{figure}

With further decreasing of the coupling strength $\sigma$, the number of solitary oscillators continues growing up, and their basin of attraction also enlarges. In this case the basin prepared for only one oscillator (like in Fig.~\ref{fig:L_SS_many}d) depends on the choice of an ensemble element. This is caused by a random distribution of solitary elements in the ensemble. In some cases the basins demonstrate a fractal structure (like in Fig.~\ref{fig:L_incoherence}d).

Finally the number of solitary oscillators is essentially the same as the number of typical oscillators (Fig.~\ref{fig:L_incoherence}). Now they can not be called as solitary oscillators. Both sets (the typical and the ``solitary'' ones) have four parts and approach each other (Fig.~\ref{fig:L_incoherence}). For each solitary oscillator the basin of attraction to the solitary set becomes fractal (Fig.~\ref{fig:L_incoherence}). This is the transition to chaotic spatial incoherence.

\begin{figure}[htbp]
\centering
\resizebox{1\columnwidth}{!}{\includegraphics{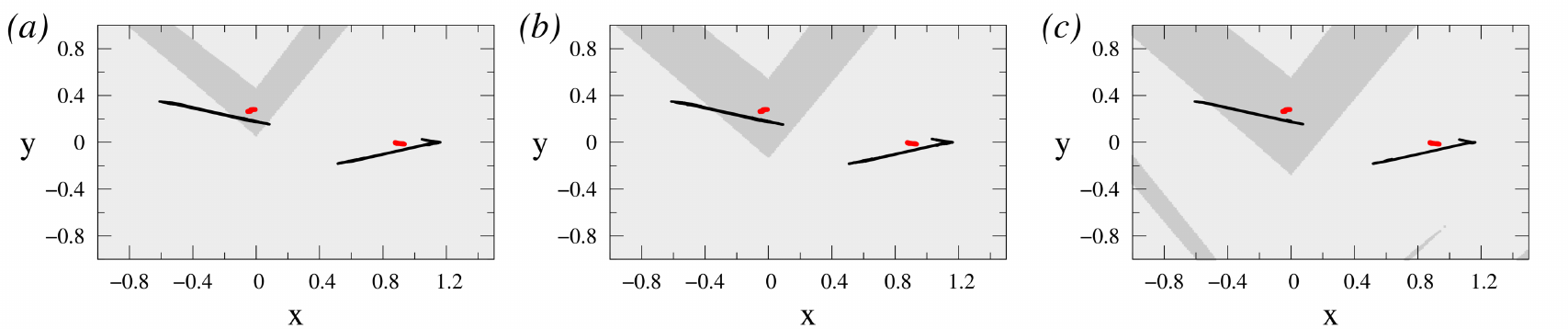} }
\caption{Basins of attraction for a group of oscillators in the solitary state regime for different numbers of oscillators in the group: (a) 10; (b) 200; (c) 400. Parameters: $\alpha=1.4$, $\beta=0.3$, $\sigma=0.22$ and $r=0.2$.}
\label{fig:special_basins}
\end{figure}

To illustrate the effect of specially prepared initial conditions let us consider the regime of several solitary states for $\sigma=0.22$ and $r=0.2$. We now specify identical initial conditions for a certain number of oscillators and explore how in their basins of attraction change (see Fig.~\ref{fig:special_basins}a--c).Thus, one can enlarge their basin of attraction by increasing the number of solitary oscillators. It means that the principle ``the wider basin, the larger number of solitary states'' works in both sides.

\begin{figure}[htbp]
\centering
\resizebox{1\columnwidth}{!}{\includegraphics{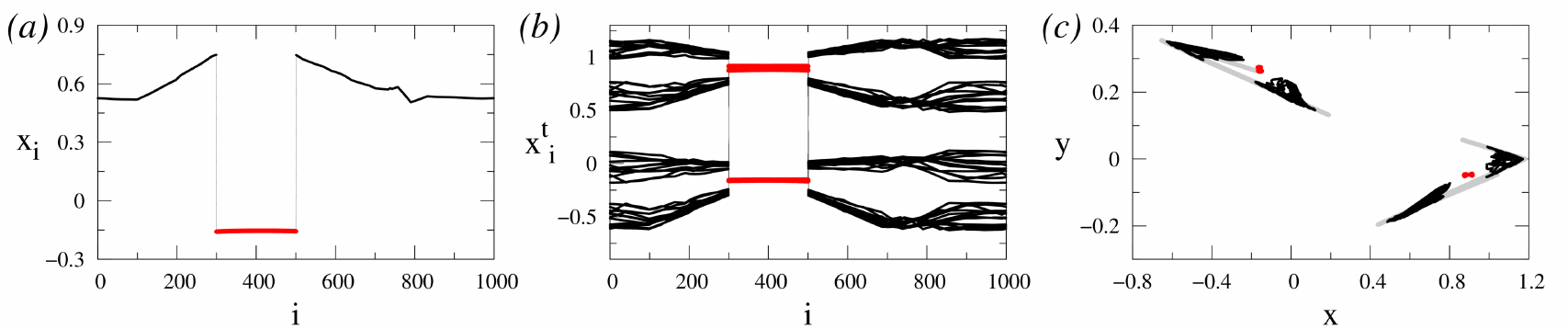} }
\caption{Specially prepared cluster of oscillators in the solitary state mode. Panel (a) corresponds to the instantaneous snapshot at $t=10000$, panel (b) depicts the last 50 instantaneous spatial profiles $x^t_i$ and the panel (c) illustrates the coexistence of typical (black) and solitary (red) sets in the ensemble (\ref{eq:system}) of Lozi maps. Parameters: $\alpha=1.4$, $\beta=0.3$, $\sigma=0.226$ and $r=0.2$.}
\label{fig:special_snapshots}
\end{figure}

Using specially prepared initial conditions for a group of oscillators on can create a cluster of oscillators in this regime, as it is shown in Fig.~\ref{fig:special_snapshots}. All the parameters are the same as in Fig.~\ref{fig:L_SS1} for the random initial conditions. This effect confirms the existence of the second attractor and the fact that solitary state is caused by the multistability of individual elements in the system (\ref{eq:system}).

\section{Modelling the individual element dynamics in the ensemble (\ref{eq:system}) using a nonautonomous map}\label{sec:simplification}

Let us consider a single element (node) of the ensemble (\ref{eq:system}) which is focused by the neighbouring oscillators. We choose the oscillator $i=k$ and denote $x^t_k=x^t$, $y^t_k=y^t$. This enables us to rewrite the equation for the $k$th oscillator in the following form:
\begin{equation}\label{eq:map_simplification}
x^{t+1}=(1-\sigma) f_x(x^t, y^t) + F^t, \ \ \ \ \ \ \ \ \ y^{t+1}=f_y(x^t, y^t),
\end{equation}
where $F^t=\frac{\sigma}{2P} \times \sum\limits^{j=i+P}_{j=i-P} f(x^t_j,y^t_j),\ \ j\neq i$.

Equation (\ref{eq:map_simplification}) indicates that the coefficient $\sigma$ affects the effective parameters of individual elements. In fact, if the term $F^t$ is discarded, the equation (\ref{eq:map_simplification}) can be transformed to the form (\ref{eq:system}) by letting $x=(1-\sigma)X$, $y=(1-\sigma)Y$:
\begin{equation}
X^{t+1}=1-\alpha_\mathrm{eff} |X^t|+X^t, \ \ \ \ Y^{t+1}=\beta_\mathrm{eff}X^t, \nonumber
\end{equation} 
where $\alpha_\mathrm{eff}=(1-\sigma)\alpha$ and $\beta_\mathrm{eff}=(1-\sigma)\beta$ are the effective values of parameters.

If the coupling radius is too large, then $F^t$ can be regarded as a mean field showing an averaging impact of neighbours on the considered element. The impact of the $k$th oscillator on the dynamics of the other elements in the ensemble may be ignored in the first approximation. Let us consider the case when all neighbouring oscillators are in the typical regime. It means that their instantaneous states belong to one part of the double-band attractor which is demonstrated by the autonomous map (\ref{eq:Lozi}) with the parameters $\alpha$ and $\beta$. It allows us to replace $F^t$ by a certain value $x^*$ averaged over all $x$ from the corresponding part of the attractor.

\begin{figure}[htbp]
\centering
\resizebox{0.8\columnwidth}{!}{\includegraphics{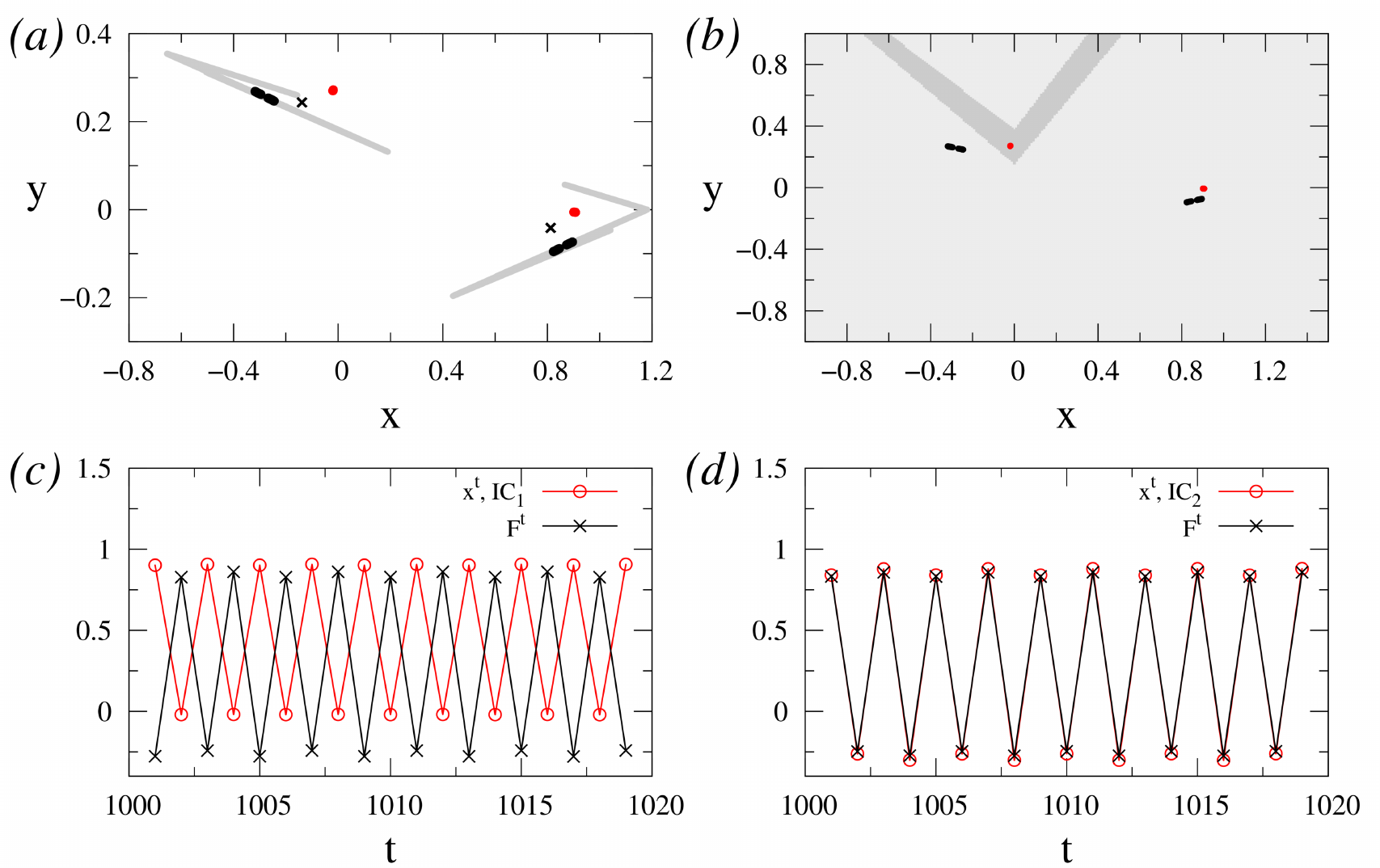} }
\caption{Simulation results for an individual element in the ensemble (\ref{eq:system}) using the system (\ref{eq:map_simplification}). The panel (a) shows the sets which can be observed in this system: light-gray circles represent to the attractor of the Lozi map (\ref{eq:Lozi}) with $\alpha=1.4$, $\beta=0.3$; x-points correspond to the autonomous system (\ref{eq:map_simplification}) ($F^t=0$); red and black points indicate coexisting sets which can be observed for different initial conditions with $\sigma=0.226$ and $F^t\neq 0$. Their basins of attraction are shown in (b). The light-gray region in the panel (b) conforms to the black set and the gray region corresponds to the red set. Panels (c) and (d) depict the external force $F^t$ and realizations of the system (\ref{eq:map_simplification}) in the typical regime (d) (black points in panel (a)) and in the ``solitary'' one (c) (red points in the panel (a)).}
\label{fig:simplification}
\end{figure}

We now explore the system (\ref{eq:map_simplification}) for the Lozi map with parameters $\alpha=1.4$, $\beta=0.3$ and $\sigma=0.226$. The ensemble (\ref{eq:system}) of Lozi maps with these parameters and $r=0.2$ demonstrates the regime with only one solitary oscillator (Fig.~\ref{fig:L_SS1}). Figure \ref{fig:simplification}a shows the attractor of the map (\ref{eq:map_simplification}) without the external force ($F^t=0$) (black X-points) and the attractor of the Lozi map (\ref{eq:Lozi}) with the effective values of parameters (light-gray circles). They are not coincide because of the multiplier $(1-\sigma)$ in the first equation (\ref{eq:map_simplification}). The chosen value $\sigma$ corresponds to the period-2 dynamics in the autonomous system (\ref{eq:Lozi}). Now we add the term $F^t$ to modulate the influence of neighbours in the ensemble. We are focused on the regime of typical state for all neighbouring oscillators. It allows us to replace their impact by averaged values of variables on one part of the Lozi attractor. Taking into account the period-4 dynamics of the typical state (see Fig.~\ref{fig:L_SS1}b) we consider every 4 value of $x^t$-realizations of the Lozi map (\ref{eq:Lozi}) and then we obtain the time-mean value.

Stable regimes of the system (\ref{eq:map_simplification}) strongly depend on initial conditions. For the most part of initial conditions one can obtain the typical set shown by blue points in Fig.~\ref{fig:simplification}a. However, some initial conditions lead to another set (red points in Fig.~\ref{fig:simplification}a). This set is almost the same as the set demonstrated by a single Lozi map (\ref{eq:Lozi}) with the effective parameters $\alpha_\mathrm{eff}$ and $\beta_\mathrm{eff}$. But in the case of the system (\ref{eq:map_simplification}) with the external force, this set consists of a large number of closely spaced points. The basins of attraction for these two coexisting sets are shown in Fig.~\ref{fig:simplification}b for the left part of the blue set and the right part of the red set. It can be seen that the red set has a V-form of the basin. The same form has been obtained for solitary states (Fig.~\ref{fig:L_SS1}d) in the ensemble (\ref{eq:system}).

What is the mechanism of occurrence of two stable regimes in the nonautonomous oscillator (\ref{eq:map_simplification})? To answer this question, it is enough to compare their realizations with the external force wave form. The corresponding curves are shown in Fig.~\ref{fig:simplification}c,d. It is obvious that both regimes in the system (\ref{eq:map_simplification}) represent in-phase and anti-phase synchronizations of self-sustained oscillations (periodic or almost periodic) by the external force, which is also periodic. In contrast to the in-phase regime, the anti-phase mode has a narrow basin of attraction ($\mathrm{IC_1}$). The same results have been obtained for the ensemble (\ref{eq:system}) (see Fig.~\ref{fig:L_SS1}e,f). The oscillations of the solitary element $k=702$ are anti-phase with respect to other oscillators from the typical regime and the coupling term $\frac{\sigma}{2P} \times \sum\limits^{j=i+P}_{j=i-P} f(x^t_j,y^t_j),\ \ j\neq i$. The typical oscillators represent in-phase synchronization (Fig.~\ref{fig:L_SS1}f).

Thus, the model (\ref{eq:map_simplification}) enables us to understand the mechanism of appearance of solitary states in the ensemble of Lozi maps. The coexistence of two sets and their location above and below the attractor of the autonomous map (\ref{eq:Lozi}) harmonize with the results for the ensemble (\ref{eq:system}) (compare Figs. \ref{fig:L_SS1} and \ref{fig:simplification}). On the other hand, there is no quantitative correspondence between these two systems. It is caused by the neglect impact of the considered oscillator on the external force. For the limiting cases of large and small values of the coupling strength, the model (\ref{eq:map_simplification}) looses its correlation with the ensemble dynamics. The system (\ref{eq:map_simplification}) does not take into account synchronization between elements for strong $\sigma$ and the impact of the other solitary states for weak coupling $\sigma$.

The mechanism described above is realized if the following conditions are met: 1) The force $F^t$ has to be periodic or has to possess a periodic component. 2) The periodic force must lead to coexisting regimes of in-phase and anti-phase synchronization obtained for different initial conditions. In this case oscillations have to exhibit also a periodic component; 3) The induced regime of anti-phase synchronization should have a narrow basin of attraction. It leads to a small number of initial conditions which provides this set. All these conditions are fulfilled takes place in the ensemble (\ref{eq:system}) of Lozi maps in contrast to a similar ensemble of H{\'e}non maps.

\section*{Conclusions}\label{sec:conclusion}

The numerical results presented above describe one of the possible mechanisms for solitary state formation in an ensemble of nonlocally coupled oscillators. As we have anticipated, it differs from the mechanism proposed in \cite{JAR18}. It is caused by a principally different dynamics of individual elements. We have studied the ensemble of Lozi maps which demonstrate a quasihyperbolic chaotic attractor. It has been shown that in a wide range of the coupling parameter, the partial elements in the ensemble demonstrates dynamics with a clearly produced periodic component. The total impact of neighbouring oscillators also has this component. As a result, one can obtain either in-phase or anti-phase synchronization for different initial conditions. The last one has a narrow basin of attraction and corresponds to the solitary state regime. The other oscillators, which are in-phase synchronized, belong to the typical state.

We have shown that the changes in the local dynamics of an individual oscillator are caused by an almost periodic external influence. This means that all neighbours belong to the same part of the Lozi attractor.

The existence of solitary state is caused by arising of multistability in the system. The second attracting set appears for some values of parameters. In this case oscillators are attracted to different sets depending on initial conditions.

We suppose that there are different ways of the solitary state appearance. We have described the mechanism for the ensemble of chaotic maps. However, the peculiarities of the mechanism must depend on a type of individual elements and their interplay.

\section*{Acknowledgements}
This work was supported by the Russian Ministry of Education and Science (Project Code 3.8616.2017/8.9) and by the Russian Science Foundation (Grant No. 16-12-10175).


\end{document}